\documentstyle[preprint,aps,psfig]{revtex}

\tighten
\begin{document}
\draft
\preprint{TPI-MINN-99/47 $\;\;$
          UMN-TH-1822-99 $\;\;$
          DAMTP-1999-139}

\newcommand{\nc}{\newcommand}
\nc{\vivi}{very interesting and very important}
\nc{\al}{\alpha}
\nc{\ga}{\gamma}
\nc{\de}{\delta}
\nc{\ep}{\epsilon}
\nc{\ze}{\zeta}
\nc{\et}{\eta}
\renewcommand{\th}{\theta}
\nc{\Th}{\Theta}
\nc{\ka}{\kappa}
\nc{\la}{\lambda}
\nc{\rh}{\rho}
\nc{\si}{\sigma}
\nc{\ta}{\tau}
\nc{\up}{\upsilon}
\nc{\ph}{\phi}
\nc{\ch}{\chi}
\nc{\ps}{\psi}
\nc{\om}{\omega}
\nc{\Ga}{\Gamma}
\nc{\De}{\Delta}
\nc{\La}{\Lambda}
\nc{\Si}{\Sigma}
\nc{\Up}{\Upsilon}
\nc{\Ph}{\Phi}
\nc{\Ps}{\Psi}
\nc{\Om}{\Omega}
\nc{\ptl}{\partial}
\nc{\del}{\nabla}
\nc{\be}{\begin{equation}}
\nc{\ee}{\end{equation}}
\nc{\bea}{\begin{eqnarray}}
\nc{\eea}{\end{eqnarray}}
\nc{\ov}{\overline}
\nc{\gsl}{\!\not}
\newcommand{\s}{\mbox{$\sigma$}}
\newcommand{\bi}[1]{\bibitem{#1}}
\newcommand{\fr}[2]{\frac{#1}{#2}}
\newcommand{\gm}{\mbox{$\gamma_{\mu}$}}
\newcommand{\gn}{\mbox{$\gamma_{\nu}$}}
\newcommand{\Le}{\mbox{$\fr{1+\gamma_5}{2}$}}
\newcommand{\R}{\mbox{$\fr{1-\gamma_5}{2}$}}
\newcommand{\GD}{\mbox{$\tilde{G}$}}
\newcommand{\gf}{\mbox{$\gamma_{5}$}}
\newcommand{\Ima}{\mbox{Im}}
\newcommand{\Rea}{\mbox{Re}}
\newcommand{\Tr}{\mbox{Tr}}
\newcommand{\psl}{\slash{\!\!\!p}}
\newcommand{\cp}{\;\;\slash{\!\!\!\!\!\!\rm CP}}
\newcommand{\qq}{\langle \ov{q}q\rangle}

\title{Hadron Electric Dipole Moments from CP-Odd Operators of Dimension Five
 Via QCD Sum Rules: The Vector Meson}

\author{Maxim Pospelov$^1$\footnote{pospelov@mnhepw.hep.umn.edu} 
          and Adam Ritz$^{12}$\footnote{a.ritz@damtp.cam.ac.uk}\\}

\address{$^1$Theoretical Physics Institute, School of Physics and Astronomy \\
         University of Minnesota, 116 Church St., Minneapolis, MN
         55455, USA\\ $\;$\\
         $^2$\footnote{Present address}
         Department of Applied Mathematics and Theoretical Physics \\
         University of Cambridge, Silver St., Cambridge CB3 9EW, 
         UK}
\date{\today}

\maketitle

\begin{abstract}
We present a complete analysis of 
the electric dipole moment (EDM) of the
$\rho$-meson induced by CP violating operators of dimension 4 and 5 within
the QCD sum rules approach. The set of CP-odd operators includes the theta
term and the electric and chromoelectric dipole moments of the three
light quarks. We find that the $\rho$-meson EDM induced purely by the 
EDMs of quarks is smaller, but still in 
reasonable agreement, with the predictions of a
naive constituent quark model. However, the chromoelectric dipole 
moments, including that of the strange quark, give comparable and 
sometimes larger contributions. We also consider the effect on  
the hadronic EDM of the existence of Peccei-Quinn symmetry. 
When this symmetry is active, chromoelectric dipole moments induce 
a linear term in the axion potential which leads to a numerically 
important vacuum contribution to the hadronic EDM. 

\end{abstract}

\vfill\eject

\section{Introduction}

Tests of time-reversal symmetry at low energies are  
an important source of information about the CP properties 
of the physics at and above the electroweak scale, 
complementary to that coming from $K$ and $B$ meson physics. 
Impressive experimental progress achieved during the last decade has brought 
the limits on the electric dipole moments (EDMs) of 
neutrons, heavy atoms, and molecules \cite{EDMS}
down to a remarkably low level. 
The Kobayashi-Maskawa model, so successful in explaining the observed
CP violation in K mesons, predicts EDMs to be several orders of 
magnitude smaller than the current experimental sensitivity. This
presents a unique opportunity for limiting extra sources of CP-violation, and
the constraints resulting from EDM data are generally very strong \cite{KL}.

Generically, EDMs can be used to probe the physics at a
high energy scale by limiting the coefficients of operators
${\cal O}_i$ with dimension $k\geq 4$ in the effective low energy Lagrangian. 
The effective Lagrangian for these operators has the form,
\be
 {\cal L}_{eff} \sim \sum_i c_i M^{4-k}{\cal O}_i^{(k)},
\ee
where $M$ is the mass scale at which these effective operators are induced
and $c_i$ their coefficients which, in general, 
have logarithmic scale dependence. 
These operators are odd under CP transformations and 
their coefficients $c_i$ are proportional to the fundamental 
CP-violating phases of the underlying theory. 
Consequently, the calculation of a hadronic EDM can naturally be
separated into two main parts. Firstly, there is the calculation 
of the coefficients $c_i$ for a specific model of CP-violation 
which involves integrating out distances shorter than $M^{-1}$. 
The second part, which is by far the more complicated, is 
the problem of switching from the perturbative
quark-gluon description to the level of hadrons which requires
nonperturbative input. 

In this letter, we 
present a systematic analysis of the EDM of the simplest hadron
-- the $\rho$-meson (more specifically $\rh^+$) -- induced by 
operators with dimension 4 
and 5 within QCD sum rules which up to now remains the most reliable 
analytical method for calculating the properties of hadrons \cite{sr}.
We choose the following parametrization of 
dimension four and five CP violating sources:
\be
 \de {\cal L} = \bar\th \frac{\al_s}{8\pi}G\tilde{G}
               -\frac{i}{2}\sum_{q=u,d,s}d_q \ov{q}F\si \ga_5 q
               -\frac{i}{2}\sum_{q=u,d,s}\tilde{d}_q \ov{q}g_sG\si \ga_5 q,
   \label{deL}
\ee
where $F_{\mu\nu}$ and $G_{\mu\nu}$ are the electromagnetic and
gluonic field strength tensors.
The first term here is the effective theta term, which primarily is
due to the fundamental QCD vacuum angle $\th_{QCD}$.
This represents a challenge for particle physics model building 
as naturally one would expect $\bar\theta\sim  O(1)$ which is ruled 
out by all available data on EDMs. As a consequence,
one is usually led to introduce the Peccei-Quinn (PQ) mechanism via
which this primary source of $\bar\th$ is removed, $\th_{QCD}=0$.
Nonetheless, even in the presence of PQ symmetry, as we shall
discuss, the effective $\th$-term is non-vanishing.
The two sums in (\ref{deL}) represent the EDMs and chromoelectric
dipole moments (CEDMs) of the three light quark flavors. 
Other quarks are considered to be heavy and can thus be integrated out 
producing operators of dimension 6 and higher. 

To motivate this calculation, we note that 
in principle experiments \cite{EDMS}
impose strong constraints on some combination of 
the coefficients $\bar\theta,\; d_i$, and $\tilde d_i$ and thus on the 
fundamental CP-phases of the theory. In order to extract these 
constraints, however, we have to embark on a non-perturbative 
calculation of the electric dipole moment of the neutron in particular 
(or nucleon-nucleon interaction, as in the case of the EDM of $^{199}$Hg) 
induced by the effective Lagrangian (\ref{deL}).
The connection between different EDM observables and the coefficients 
$\bar\theta,\; d_i, \; \tilde d_i$
is especially important in the framework 
of supersymmetric theories where eq. (\ref{deL})
and an additional three-gluon CP-odd operator \cite{wein}
represent a complete set of the relevant operators with the 
coefficients explicitly calculable as functions of 
the soft-breaking parameters. (For a recent discussion 
in the context of the MSSM, see e.g. \cite{FOPR}).

Generically, the EDM of a hadron can be written as a linear combination
of the coefficients in eq. (\ref{deL})
\be
 d_{h} = d_{h}(\bar\th)+d_{h}^{\rm EDM}(d_u,d_d,d_s) 
    + d_{h}^{\rm CEDM}(\tilde{d}_u,\tilde{d}_d,\tilde{d}_s). \label{drho}
\ee

In the case of the neutron EDM, $d_n(\bar\theta)$ has been calculated
using various different techniques: (1) making use of the 
dominance of a pion loop-induced logarithm in the 
chiral limit \cite{CDVW}; (2) in the Skyrme model \cite{dixon}; and
finally using QCD sum rules \cite{pr2,Henley}; and all have 
produced similar results.
For the calculation of $d_{n}^{\rm EDM}(d_u,d_d,d_s)$, 
the quark tensor charges over the nucleon
are required, and various techniques \cite{ADHK,MP} have produced results
consistent with the predictions of a naive SU(6) quark model.
We note that the contribution of the strange quark 
EDM is found to be consistent with zero in all methods. 
 
Unfortunately, the quantitative evaluation of the CEDM contribution to 
the neutron EDM is considerably more complicated and although
a number of serious attempts have been made to 
estimate $d_{n}^{\rm CEDM}(\tilde{d}_u,\tilde{d}_d)$, these results
often differ by more than one order of magnitude. 
In particular, direct QCD sum rules calculations \cite{KKY,kw} 
give $d_n$ $\sim 20$ times smaller than 
the estimates based on the chiral loop approach \cite{KKZ,KK}. 
Thus, we believe that an independent calculation of 
$d_{n}^{\rm CEDM}(\tilde{d}_u,\tilde{d}_d,\tilde{d}_s)$ within 
the QCD sum rule approach is absolutely necessary in order
to clarify the magnitude and sign of the contribution of 
quark CEDMs to the EDM of the neutron. 

With this motivation in mind, the discussion of the $\rh$-meson
presented here\footnote{Note that the $\rh$-EDM was also studied
within the context of a quark model \cite{McK}.} 
will serve a dual purpose. Firstly, the detailed
analysis of the contribution of dimension four and five CP-odd  
operators to a particular hadronic EDM allows us insight into the
relative sizes of the contributions, and in particular the dominant
mechanism inducing the EDM, which may well prove quite
universal. Secondly, the $\rho$-meson is the simplest light quark system
where QCD sum rules are known to work well. Thus it is a convenient
arena in which to develop techniques that should later prove useful 
in an analogous, but much more involved, study of nucleon
sum rules and the neutron EDM. In this regard, we recently calculated 
$d_\rho(\theta)$ by generalizing the operator 
product expansion (OPE) to the case of an external $\theta$ background
\cite{pr1}, which was subsequently applied to the case 
of $d_n(\theta)$ \cite{pr2}.  In this letter this method is 
developed further to include all CP-odd dimension 5 operators. 
As a result we obtain $d_\rho$ as an explicit 
function of $\bar\theta,\; d_i$, and $\tilde d_i$. We observe 
the numerical importance of all CEDM contributions and a strong dependence
of the result on the existence/absence of PQ symmetry. With the use
of the techniques developed here and in our earlier work, 
the question of $d_{n}^{\rm CEDM}(\tilde{d}_u,\tilde{d}_d,\tilde{d}_s)$ 
will be readdressed in the near future \cite{PRprog}.

\section{OPE Analysis}

Within the sum rule approach to the $\rh^+$ EDM, 
we need to consider the correlator of currents with $\rh^+$ quantum numbers,
in a background with a CP violating source and an electromagnetic 
field $F_{\mu\nu}$,
\be
 \Pi_{\mu\nu}(-p^2) = i\int d^4x e^{ip\cdot x}
    \langle 0|T\{j^+_{\mu}(x)j^-_{\nu}(0)\}|0\rangle_{\cp,F}.
\ee
In the presence of these sources, and since we will work outside 
the chiral limit with unequal quark masses,
it is necessary to take into account mixing between the vector 
current associated with $\rh^+$ and
the axial-vector current \cite{pr1}. 
Thus we parametrize the full current in form
\be
 j^{+}_{\mu} = V_{\mu} +icA_{\mu},
\ee 
where $c$ is a (real) mixing parameter to be determined, and
\be
 V_{\mu}=\ov{u}\ga_{\mu}d\;\;\;\;\;\;\;A_{\mu}=\ov{u}\ga_{\mu}\ga_5d.
\ee
The parameter $c$ is linear in the CP violating source, and thus
to first order we have,
\be
 \Pi_{\mu\nu} = \langle V_{\mu}V_{\nu}^{\dagger}\rangle_{\cp}
           +ic(\langle V_{\mu}A_{\nu}^{\dagger}\rangle
          -\langle A_{\mu}V_{\nu}^{\dagger}\rangle)+\cdots
    \label{fullPi}
\ee
In evaluating these correlators, we require 
the propagator in the presence of CP violating sources, 
$S^q_{ab}\equiv \langle q_a(x) \ov{q}_b(0)\rangle_{\cp,F}$,
which has the form
\be
\int d^4x e^{ip\cdot x} S^q_{ab}(x) = i\frac{(\psl + m{\rm\bf 1})_{ab}}{p^2}
  -\frac{d_q}{2}\frac{(\psl F\si\ga_5\psl)_{ab}}{p^4}
   -\frac{\tilde{d}_q}{2}\frac{(\psl g G\si\ga_5\psl)_{ab}}{p^4}.
\label{prop}
\ee
 Using eq. (\ref{prop}), we can express the relevant current correlators as
\be
\int d^4x e^{ip\cdot x} \langle 0|V_{\mu}(x)V^{\dagger}_{\nu}(0)|0\rangle =  
   -i\int d^4x e^{ip\cdot x} {\rm Tr}(\ga_{\mu}S^d(x)\ga_{\nu}
    S^u(-x)),
\ee
with similar expressions for $\langle V A^{\dagger}\rangle$ 
and $\langle AV^{\dagger}\rangle$.
The appropriate tensor structure to isolate is 
$\tilde{F}_{\mu\nu}(=\ep_{\mu\nu\al\beta}F^{\al\beta}/2)$, and we
evaluate the correlators as usual by projecting to the vacuum in order
to extract this structure (see e.g \cite{pr1} for more details 
in the case of $\rh^+$). In doing so we will assume a constant
background electromagnetic field, and use a fixed point
gauge \cite{smilga} for the gluon field. Condensates involving quark
and gluon fields are then parametrized in terms of 
certain ``condensate susceptibilities''
following Ioffe and Smilga \cite{is},
\bea
 g_s\langle 0 | \ov{q} G\si q | 0 \rangle & = & -m_0^2 
    \langle 0| \ov{q}q|0\rangle \nonumber\\
 \langle 0| \ov{q}\si_{\mu\nu}q|0\rangle_F & = & e_q\ch F_{\mu\nu}
           \langle 0| \ov{q}q|0\rangle \nonumber\\
 g_s\langle 0| \ov{q}(G_{\mu\nu}^at^a)q|0\rangle_F & = & e_q\ka F_{\mu\nu}
           \langle 0| \ov{q}q|0\rangle \nonumber\\
 2g_s\langle 0| \ov{q}\ga_5(\tilde{G}_{\mu\nu}^at^a)q|0\rangle_F 
   & = & ie_q\xi F_{\mu\nu}\langle 0| \ov{q}q|0\rangle .
\eea

%%%%%%%%%%%%%%%%%%%%%%%
\begin{figure}
 \centerline{%
   \psfig{file=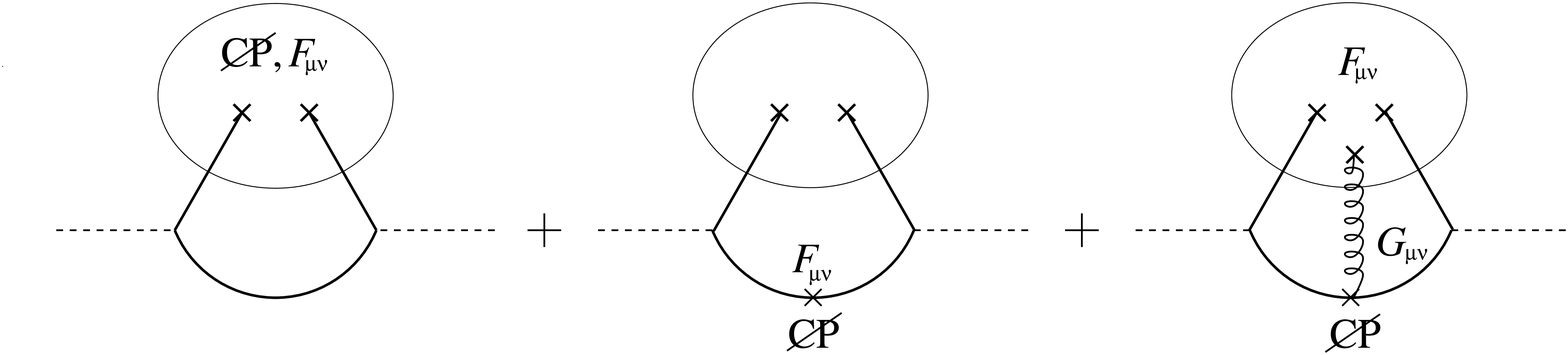,width=12cm,angle=0}%
         }
 \caption{Contributions to the correlator at leading order in $F_{\mu\nu}$.}
\end{figure}
%%%%%%%%%%%%%%%%%%%%%%%
\noindent
In general, the OPE will contain perturbative pieces in addition to terms 
explicitly proportional to the quark-gluon condensates.
However, it turns out that the simple perturbative 
contributions are heavily suppressed. This is due to the 
chirality-flipping structure of the EDM and CEDM operators. The result
then vanishes unless there is an additional chirality flip  
in the loop which can only come from a quark mass in the propagator.
Indeed, for the $\theta$-induced 
contribution, the perturbative piece is identically 
zero \cite{pr1}.

Thus, we are left with only the non-perturbative contributions, for which 
the leading order diagrams are given in Fig. 1. 
Naively, one could drop all the terms in the quark propagator
which are proportional to a small mass, $m_u$ or $m_d$. 
This is not valid, however, as it turns out that at different 
stages of the OPE calculation, quark masses from the propagator,
$m_{u(d)}$, enter in front of vacuum correlators of the form 
$C \sim \int d^4 x\langle 0|T(\ov{u}\ga_5 u, \bar u (G\s)\gf u(x)|0 \rangle$. 
These correlators can be saturated by pion exchange, producing a 
non-vanishing contribution in the chiral limit, $m_q C \sim O(m_q/m_\pi^2)
\rightarrow $ const. Equivalently, one can perform a chiral rotation, requiring
that pions cannot be produced from the vacuum, 
$\langle 0|{\cal L}_{eff}|\pi_0 \rangle=0$. This leads to the appearance
of additional $\gamma_5$-mass terms in the Lagrangian which 
will certainly contribute to $d_\rho$. We prefer, however, not 
to perform the chiral rotation and instead to 
account for the vacuum contributions explicitly.
 
Using this approach, we find for $\langle VV^{\dagger}\rangle$,
\bea
 \langle V_{\mu}(p)V^{\dagger}_{\nu}(0)\rangle  & = &
   \frac{i\tilde{F}_{\mu\nu}}{p^2}
   \left(i\ch\left(m_de_u\langle \ov{u}\ga_5 u\rangle_{\cp}-m_ue_d
     \langle \ov{d}\ga_5d\rangle_{\cp}\right)\right. \nonumber \\
  & & \;\;\;\;\;+\left.(d_d-d_u)\qq -\left(\ka-\frac{\xi}{2}\right)
     (e_d\tilde{d}_u-e_u\tilde{d}_d)\qq\right), 
    \label{vv}
\eea
where the first term arises from Fig.~1(a), the second from Fig.~1(b), and
the third from Fig.~1(c). A useful simplification in these calculations
follows by noting that, although naively contributing to the structure
$\tilde{F}_{\mu\nu}$ via the equations of motion,  contributions arising
from a Taylor expansion of the quark wave function, actually vanish! Thus
we find the rather compact expression exhibited in Eq.~(\ref{vv}).

The necessity for additional contributions, due to mixing with the 
axial vector current, now becomes clear. Indeed, if eq. (\ref{vv})
were the entire answer, the first line calculated in the external 
$\theta$ background would produce an incorrect quark mass dependence, 
differing from $m_um_d/(m_u+m_d)$. Thus its clear that we
need to address mixing with the axial current \cite{pr1}. The mixing 
parameter $c$ may be obtained to leading order by considering the
correlators $\langle VV^{\dagger}-AA^{\dagger}\rangle$, and 
$\langle VA^{\dagger}-AV^{\dagger}\rangle$ with the external field turned
off. At this order we may diagonalize on the tensor structure
$g_{\mu\nu}p^2$, and we obtain
\be
 c = \frac{1}{2}\left[\frac{i}{6}(\tilde{d}_d-\tilde{d}_u)
    \frac{m_0^2}{m_u+m_d}
   - i\frac{m_d\langle \ov{u}\ga_5 u\rangle_{\cp}-m_u\langle \ov{d}\ga_5 d
    \rangle_{\cp}}{(m_u+m_d)\qq}\right]. \label{cval}
\ee
This value of $c$ ensures the absence of mixing to leading order
between the ``eigen-currents'' $V_\mu+icA_\mu$
and $iA_\mu+cV_\mu$.
The calculation of the mixed correlators is straightforward, and the 
result is given by
\be
 \langle VA^{\dagger}\rangle = - \langle AV^{\dagger}\rangle 
     = -\frac{\tilde{F}_{\mu\nu}}{p^2}\ch(m_ue_d+m_de_u).
\ee
Combining the relevant pieces according to eq.~(\ref{fullPi}) 
we arrive at the following result: 
\begin{eqnarray}
\Pi_{\mu\nu} = 
\frac{i\tilde{F}_{\mu\nu}}{p^2}\qq\left[d_d-d_u+\chi(e_d-e_u)\fr{m_um_d}
{m_u+m_d}\fr{\langle \bar d\gf d + \bar u\gf u\rangle_{\cp}}{\qq}+
\right.\nonumber \\ \left.
\chi m_0^2(\tilde d_d - \tilde d_u)\fr{e_um_d+e_dm_u}{m_u+m_d}+
\left(\kappa-{\xi\over 2}\right)(\tilde d_d e_u - \tilde d_u e_d)\right]
\label{before}
\end{eqnarray}

The next problem to address is the calculation of the vacuum 
matrix elements in eq. (\ref{before}). 
These terms, arising from Fig.~1(a), require the evaluation of correlators
of the form $\int d^4y\langle \ov{q}\ga_5 q(x), i\de{\cal L}(y)\rangle$, where
$\de{\cal L}$ given in Eq.~(\ref{deL}) involves 
in particular the $\th$-term and 
the colour EDM sources which may be extracted from the vacuum at leading
order in the background electromagnetic field. The case of $\theta$ was 
discussed at length in our previous paper \cite{pr1}. Here we shall 
concentrate on the CEDMs. 
We will evaluate these correlators by inserting a complete set of
intermediate states $\{\ep_i\}$,
\be
 \langle\ov{q}\ga_5 q(x), i\de{\cal L}(y)\rangle = \langle\ov{q}\ga_5 q(x)
    \left(\sum_i |\ep_i\rangle \frac{i}{-m_{i}^2}\langle \ep_i |\right)
   i\de{\cal L}(y)\rangle.
\ee
 For the two-flavour case, to a good approximation
we need only consider $\pi_0$, while for three flavours we also include
$\et$. The neglected states then have masses $\geq 1$GeV and will remain 
massive in the chiral limit. They
may contribute corrections of no more than 25\%, which we shall factor
into our precision estimate. The remaining condensates may be
reduced to commutators in a manner analogous to the soft pion theorem in
chiral perturbation theory. For a generic operator ${\cal O}$ we have
\bea
 \langle 0| {\cal O} | \pi_0 \rangle & = & \frac{i}{2f_{\pi}}
       \langle 0| [{\cal O}, u^{\dagger}\ga_5u-d^{\dagger}\ga_5d]|0\rangle\\
  \langle 0| {\cal O} | \et \rangle & = & \frac{i}{2\sqrt{3}f_{\pi}}
       \langle 0| [{\cal O}, u^{\dagger}\ga_5u+d^{\dagger}\ga_5d
      -2s^{\dagger}\ga_5 s]|0\rangle.
\eea
Concentrating on the colour EDM sources in (\ref{deL}) we then obtain
\bea
 \langle \ov{u}\ga_5 u \rangle_{\pi_0} 
    &=& -\langle \ov{d}\ga_5 d\rangle_{\pi_0} = \frac{i}{2}
      (\tilde{d}_u-\tilde{d}_d)\qq
      \frac{m_0^2}{m_u+m_d} \\
  \langle \ov{u}\ga_5 u \rangle_{\et} 
    &=& \langle \ov{d}\ga_5 d\rangle_{\et} = -\frac{i}{4}\tilde{d}_s\qq 
      \frac{m_0^2}{m_s} +\cdots,  
\eea
where in the last expression we have neglected terms of 
$O(\tilde d_{(u,d)}/m_s)$, assuming an approximate proportionality 
of the CEDMs to the quark masses, i.e.
$\tilde d_d /\tilde d_s \sim m_d/m_s\ll 1$.

Putting the pieces together, and decomposing into singlet and triplet 
combinations, we finally obtain the following
result for the correlator in (\ref{fullPi}),
\bea
 \Pi_{\mu\nu} & = & \frac{i\tilde{F}_{\mu\nu}}{p^2}\qq\left[
                -\fr{\chi}{2}e_-m_*\bar\theta+d_- \right. \nonumber\\
   & & \;\;\; +\left.
   \frac{\ch m_0^2}{12}\tilde{d}_-(e_+-e_-\tilde{m})
    +\frac{\ch m_0^2}{4}\tilde{d}_s e_-
    \frac{m_*}{m_s}+\frac{1}{2}\left(\ka-\frac{\xi}{2}\right)
     (e_-\tilde{d}_+-e_+\tilde{d}_-)
    \right],  \label{OPEres}
\eea 
where we have introduced the notation: $d_{\pm} = d_u \pm d_d$,
$\tilde{d}_{\pm}=\tilde{d}_u\pm \tilde{d}_{d}$,
$e_{\pm}=e_u \pm e_d$, $m_*=m_um_d/(m_u+m_d)$, and 
$\tilde{m}=(m_u-m_d)/(m_u+m_d)$.

\section{The EDM and Peccei-Quinn Symmetry}

The phenomenological side of the sum rule for the correlator $\Pi_{\mu\nu}$
may be obtained from the form-factor Lagrangian,
${\cal L} = -2id_{\rh}m_{\rh}\rh^+_{\mu}\tilde{F}_{\mu\nu}\rh^+_{\nu}
+\cdots$. We find,
\be
 \Pi_{\mu\nu}^{\rm Phen} = -2id_{\rh}m_{\rh}\tilde{F}_{\mu\nu}
    \frac{\la^2}{(p^2-m_{\rh}^2)^2}+\cdots,
\ee
where $\la$ is the coupling to $\rh^+$, and we have kept only
the double pole term. We now follow standard practice, equating
this result with (\ref{OPEres}) and performing a Borel transform to 
suppress the contribution of continuum states. Rather than
presenting a detailed analysis of the sum rule, including a
parameterization of single pole and continuum states, we shall
consider only the double pole term here, as previous
work \cite{pr1,pr2} indicates that the corrections which
arise from a more careful analysis are overwhelmed by errors from
other sources. At this order, the coupling may be obtained in terms of
the Borel mass $M$
as $\la^2=m_{\rh}^2M^2/4\pi^2$ 
from the CP even sum rule (see e.g.
\cite{rry}). It turns out that within these approximations
$d_\rho$ does not depend on $M^2$.
We then obtain the following result for the 
EDM induced by $\ov{\th}$, $(d_u,d_d)$ and the colour sources 
$(\tilde{d}_u,\tilde{d}_d,\tilde{d}_s)$,
\bea
 d^{{\rm EDM+CEDM}+\th}_{\rh} & = & \frac{2\pi^2}{m_{\rh}^3}(-\qq)\left[
 -\fr{\ch e_-m_*\bar\theta}{2}+d_-+
   \frac{\ch m_0^2}{12}\tilde{d}_-(e_+-e_-\tilde{m})\right.\nonumber\\
     & & \;\;\;\;\;\;\left.+\frac{\ch m_0^2}{2}\tilde{d}_s e_-
    \frac{m_*}{m_s}+\frac{1}{2}\left(\ka-\frac{\xi}{2}\right)
     (e_-\tilde{d}_+-e_+\tilde{d}_-)
    \right]. \label{drho1}
\eea
This result parameterizes the effect of 
all the dimension four and five sources, including the 
$\th$-term which was considered previously in \cite{pr1}. 

The numerical values for the condensates were obtained 
in \cite{BI}, \cite{chival} and \cite{kw},
\begin{eqnarray}
 m_0^2 & = & 0.8 \mbox{ GeV}^2 \mbox{\cite{BI}}\\ 
 \ch & = & - 5.7 \pm 0.6 \mbox{ GeV}^{-2} \mbox{\cite{chival}} \\
 \ka & = & - 0.34 \pm 0.1  \mbox{\cite{kw}} \\
 \xi & = & - 0.74 \pm 0.2  \mbox{\cite{kw}}
\end{eqnarray}
Note that with these values the combination $(\kappa -\xi/2)$
actually vanishes within 
the specified precision. Substituting the 
numerical values for the condensates, masses, and charges, 
we obtain a final result for the EDM of the $\rho$-meson:
\be
d_\rho = \fr{4.4\cdot 10^{-3}}{1 {\rm GeV}} \bar\theta
         +0.51(d_u-d_d) - 0.13e (\tilde d_u -\tilde d _d) - 
\fr{3.5{\rm MeV}}{m_s}e\tilde d_s
\label{num1}
\ee
Here we have used $m_u= 4.5$ MeV, and $m_d = 9.5$ MeV. We will comment
on error estimates in Section IV.

It is interesting to observe that the central point for the 
QCD sum rule prediction for $d_\rho$ induced purely by the 
EDMs of the quarks is essentially half 
the prediction of the naive non-relativistic quark model,
\be
d_\rho^{NQM}=d_u-d_d
\ee
Nonetheless, given the relatively large error bounds on 
(\ref{num1}), and noting that
smaller values of $m_u$ and $m_d$ will bring the two results closer
together, we feel that the results are in reasonable agreement. 

The contributions of up- and down-quark  CEDMs are not, in fact, any
smaller than those of the quark EDMs. This may be seen in a  
simple example of supersymmetric CP-violation due to a squark-gluino
loop. The 
induced EDM operators will then be proportional to the charges of 
the quarks, thus bringing in an additional factor of $e/3$ 
to the contribution induced by $d_d$, leading to a result of the same
order as the CEDM contributions. 
Our result for $d_\rho$(CEDMs) is larger than one would expect 
from power-counting rules such as ``naive dimensional analysis''
\cite{GM} which are often applied to estimate $d_n$(CEDMs).  
The importance of CEDM contributions in the QCD sum rule 
approach follows directly from the values of the $\chi$ and 
especially $m_0^2$ condensates. It seems clear that the
estimate within naive dimensional analysis 
corresponds to a group of perturbative diagrams in which the 
gluon from $\tilde d_q$ is attached to a quark line. 
This would correspond to an $O(\alpha_s/(4\pi))$ suppression relative
to the leading order OPE terms considered here.

Perhaps the most interesting point to make here is the significance of the 
contribution arising from the strange quark CEDM operator. The $1/m_s$ 
suppression is in fact fictitious as we expect $d_q\sim m_q$. 
When this ansatz is assumed, the contribution of the $s$-quark CEDM 
becomes even larger than that of the down quark! 

However,
the presence of the theta term is expected to be numerically 
dominant, as it corresponds to a dimension four operator and does not 
experience any suppression by a heavy mass scale.
This poses a serious problem for any model of CP-violation other than 
Kobayashi-Maskawa, as corrections to the theta term are normally
large and need to be cancelled by extreme fine-tuning of the
``initial condition'' for theta.
Currently, the accepted recipe to avoid a ``$\th$-dominated'' EDM is 
to assume Peccei-Quinn symmetry. This mechanism is 
apparently a necessity for any SUSY model operating 
with CP-SUSY phases around the electroweak scale.
 
The  existence of this mechanism brings an additional contribution to 
the EDM, not contained in eq. (\ref{drho1}). PQ symmetry, 
although allowing the axion to
set $\th_{\rm QCD}=0$, still leads to CP violating terms due to linear
contributions to the axion potential \cite{BUP}. 
In particular, the axion potential
has the form $V \sim -\th^2K - 2\th K'$, where
\be
 K = i\left<\frac{\al_s}{8\pi}G\tilde{G},\frac{\al_s}{8\pi}G\tilde{G}\right>
\ee
is the topological susceptibility, and 
\be
 K' = i\left<\frac{\al_s}{8\pi}G\tilde{G},-\frac{i}{2}d_q\ov{q}G\si\ga_5 q
      \right>
\ee
are correlators arising from the CEDM sources. This linear
shift in the axion potential then leads to an ``induced'' $\th$-term
with coefficient
\be
 \th_{\rm ind} = -\frac{K'}{K} = \frac{m_0^2}{2}\sum_{q=u,d,s}
     \frac{d_q}{m_q}.
\ee
This result follows by evaluating the correlators in the manner
described in \cite{SVZK} (see also \cite{C}).
The value of $\th_{\rm ind}$ is, of course, independent of the
axion coupling constant, and of the particular 
manner in which PQ symmetry is implemented. 
Consequently, we find an additional vacuum contribution to
the EDM in (\ref{vv}) and (\ref{cval}) of the form,
\be
 \langle \ov{u}\ga_5 u\rangle_{\th} + \langle \ov{d}\ga_5 d \rangle_{\th}
   = i\th_{ind} \qq 
   = \frac{i}{2}m_0^2\left(\sum_{q=u,d,s}\frac{d_q}{m_q}\right)\qq.
\ee
Including this expression\footnote{Note that this contribution only
affects terms proportional to $\ch m_0^2$ as the contributions
proportional to $\xi$ and $\ka$ are subleading in the EDM induced by
$\th$ \cite{pr1}.} in (\ref{drho1}) we observe the complete 
cancelation of the term proportional to the strange quark CEDM,
while this new vacuum source of $\tilde d_u$ and $\tilde d_d$ 
is combinatorially
less suppressed than the direct contributions appearing in
(\ref{drho1}). The final result then has the form
\bea
 d^{\rm EDM+CEDM+PQ}_{\rh} & = & \frac{2\pi^2}{m_{\rh}^3}(-\qq)\left[d_-+
   \frac{\ch m_0^2}{12}\tilde{d}_-(e_+-e_-\tilde{m})\right.\nonumber\\
   & & \;\;\;\;\;\left. 
    +\frac{\ch m_0^2}{4}(\tilde{d}_+-\tilde{d}_-\tilde{m})e_-
     +\frac{1}{2}\left(\ka-\frac{\xi}{2}\right)
     (e_-\tilde{d}_+-e_+\tilde{d}_-)
    \right].
\eea
Inserting numerical values for the condensates, masses, and charges
as before, we find:
\be
 d_{\rh} = 0.51(d_u-d_d) - 0.34 e(\tilde{d}_u-\tilde{d}_d)
            - 0.58 e (\tilde{d}_u+\tilde{d}_d).
\ee
We see that in the presence of PQ symmetry, the contribution
from the CEDM sources is even more pronounced.

\section{Discussion}

We have studied the EDM of the $\rho$-meson induced by
CP-odd operators of dimension four and five. 
We find that, at leading order, QCD sum rules predict that $d_\rho$ 
induced by the quark EDMs is within a factor of two of the prediction
of the nonrelativistic quark model. Moreover, a conservative estimate
of the error, accounting for higher dimension operators, neglected
intermediate states, and single pole contributions, would suggest that
this factor of two could easily be accomodated within  
the precision of our estimates. This precision could of course
be improved by including next-to-leading order 
corrections in the OPE, and also accounting for the single pole
contributions on the phenomenological side. An important 
advantage of the QCD sum rules approach to the calculation of
hadronic EDMs over other methods is the proportionality 
of the result to $d_q\qq$ which is normalization-point independent 
as in most of the models we have $d_q\sim m_q$. The 
uncertainty related with the poor knowledge of $m_q$ for the light 
quarks is thus significantly reduced.  

We find that $d_\rho$ induced by the colour EDM operators 
is actually comparable in magnitude with $d_\rho$(EDMs). There 
is no specific mechanism of suppression which can be attributed to
$d_\rho$(CEDMs). This conclusion supports an estimate of $d_n$ 
made in Ref.~\cite{KK}. We also wish to stress the numerical 
importance of the color EDM operator of the strange quark, 
whose contribution is of the same order as that 
arising from the up and down CEDM operators. 

Peccei-Quinn symmetry, which removes $\theta\sim 1$, is an apparent 
necessity for any model with large CP-violating phases other
than Kobayashi-Maskawa. When the PQ mechanism is active, the axion vacuum 
experiences a linear shift induced by CEDMs. We find that the
contribution of the up and down quark CEDMs in this case
becomes even larger, whereas the contribution of the 
strange quark CEDM is completely canceled at this order. 
This is an important result which may well prove generic and apply
also to the experimentally relevant case of the neutron EDM.
This important problem will be addressed in a subsequent 
publication \cite{PRprog}.

\bigskip
{\bf Acknowledgments}
This work was supported in part by
the Department of Energy under Grant No. DE-FG02-94ER40823.

\bibliographystyle{prsty}

\end{document}